\documentclass[useAMS,usenatbib]{mn2e}

\usepackage{epsf}
\usepackage{lscape}
\usepackage{graphicx}
\usepackage{epsfig}
\usepackage{times}
\usepackage{natbib}
\usepackage{rotating}

\def\edit{}

\begin{document}

\title[Inverse-Compton emission from the lobes of 3C\,353]{Inverse-Compton emission from the lobes of 3C\,353}
\author[J.L. Goodger et al.]  
{J.L. Goodger,$^1$\thanks{E-mail: j.l.goodger@herts.ac.uk} M.J. Hardcastle,$^1$ J.H. Croston,$^1$ N.E. Kassim$^2$ and R.A. Perley$^3$\\
$^1$University of Hertfordshire, College Lane, Hatfield, Hertfordshire AL10 9AB, UK \\
$^2$US Naval Research Lab 4555 Overlook Ave., SW Washington, DC 20375, USA\\
$^3$National Radio Astronomy Observatory, P.O. Box O, Socorro, NM 87801, USA}

\date{}

\pagerange{\pageref{firstpage}--\pageref{lastpage}} \pubyear{2007}

\maketitle

\label{firstpage}

\begin{abstract}

X-ray emission due to inverse-Compton scattering of microwave
background photons by electrons in the lobes of powerful radio
galaxies has now been seen in a large number of objects.  Combining an
inverse-Compton model for the lobe X-ray emission with information
obtained from radio synchrotron emission provides a method of
constraining the electron population and magnetic field energy
density, which cannot be accomplished using the radio data alone.
Using six frequencies of new and archival radio data and new {\it
XMM-Newton} observations of the Fanaroff \& Riley class II radio
galaxy 3C\,353, we show that inverse-Compton emission is detected in
the radio lobes of this source at a level consistent with what is seen
in other objects.  We argue that variations in the X-ray/radio ratio
in the brighter eastern lobe require positionally varying magnetic
field strength.  We also examine the X-ray nucleus and the cluster,
Zw\,1819.1-0108, spatially and spectrally.

\end{abstract}


\section{Introduction}

Extended X-ray emission from the lobes of powerful, FR\,II
\citep{fr74jlg} radio galaxies and quasars is thought to be produced
by inverse-Compton scattering of the cosmic microwave background (CMB)
\citep{feigelson95jlg,tashiro98jlg,tashiro01jlg,isobe02jlg,hardcastle02jlg}
and of infrared photons from the core \citep*{brunetti97jlg}.
Although the synchrotron self-Compton emission from the lobes can also
be modelled, it is negligible compared to the inverse-Compton emission
from other populations except in the very smallest lobes
\citep{hardcastle02jlg}.  Recent studies of the integrated X-ray
properties of radio lobes have been carried out for large samples
\citep{croston05jlg, kataoka05jlg}, showing that the CMB is the
dominant photon population in most cases.  Combining an
inverse-Compton/CMB model for the lobe X-ray emission with information
obtained from radio synchrotron emission provides a method of
constraining the electron population and magnetic field energy
density, which cannot be accomplished using radio data alone. Unlike
the jets, the lobe material is not moving relativistically so there
are no beaming effects to consider during this analysis
\citep[e.g.][]{mackay73jlg}.  Understanding the contributions of the
electron densities of the synchrotron and inverse-Compton electron
populations and the magnetic field allows us to estimate directly the
total energy in the radio source, and thus the amount of energy that
can be transferred to the environment, without assuming equipartition,
and also allows us to investigate the distribution of internal energy
within the source.

\citet{hardcastle05jlg} recently carried out a spatially resolved
X-ray inverse-Compton study of the bright X-ray and radio source
Pictor A.  They used the variation of the X-ray/radio ratio across the
lobes to investigate the variation in electron density and magnetic
field strength throughout the source.  They found that variations in
either component alone could not explain the observed X-ray and radio
properties of the lobe.  Similar spatially resolved X-ray studies of
other bright FR\,II radio galaxies are needed to follow up these
results.  Here we report on observations of the radio galaxy 3C\,353.

3C\,353 is a FR\,II radio galaxy associated with the cluster
Zw\,1718.1-0108.  Although it is one of the brightest extragalactic
sources in the sky at low frequencies, it is relatively poorly
studied, presumably due to its low declination and low Galactic
latitude.  \citet{swain96jlg} observed 3C\,353 at four radio
frequencies with the NRAO Very Large Array (VLA) and published a study
of the polarisation variation across the jets, which favoured a
spine-sheath model \citep*{swain98jlg}. Their VLA observations
revealed filamentary structure within the lobes making 3C\,353 an
excellent target for a spatially resolved study of electron
distribution and magnetic fields.  X-ray emission from the cluster
associated with 3C\,353 was first detected by \citet{iwasawa00jlg} in
{\it ASCA} and {\it ROSAT} images.  The radio source resides
in a giant elliptical galaxy on the edge of the cluster, which Iwasawa
et al. found to be bright in the X-ray with a luminous point source
coinciding with the radio galaxy's core.  Iwasawa et al. determined
global cluster temperatures and used optical observations to identify
the cluster's members.  They confirmed a redshift of $z=0.0304$ for
3C\,353, a redshift of $z=0.028$ for three bright member galaxies, and
used the velocity distribution to confirm 3C\,353 as a member of the
same system.

In this paper we use new {\it XMM-Newton} and radio observations to
investigate the nature of the electron distribution and the magnetic
fields in the lobes and hotspots of 3C\,353.  We also determine the
cluster temperature, density and pressure profiles and examine the
cluster's interaction with the lobes of 3C\,353. Throughout the paper
we use a cosmology in which $H_{0} = 70$\,km\,s$^{-1}$\,Mpc$^{-1}$,
$\Omega_{m} = 0.3$, and $\Omega_{\Lambda} = 0.7$.  The angular scale
is 1\,arcsec = 0.61\,kpc.  We define the spectral index in the sense
that $S_{\nu} \propto \nu^{-\alpha}$.  The photon index $\Gamma = 1 +
\alpha$.

\section{Data}

We used a combination of new and archival radio data and new {\it
XMM-Newton} X-ray data to examine the broad-spectrum electron energy
distribution of 3C\,353.  Our radio data extends to lower frequencies
than have been studied previously, whilst our X-ray data has greatly
improved sensitivity and resolution compared to the {\it ASCA} data of
\citet{iwasawa00jlg}.
 
\begin{figure*}
\includegraphics[width=1.01\textwidth]{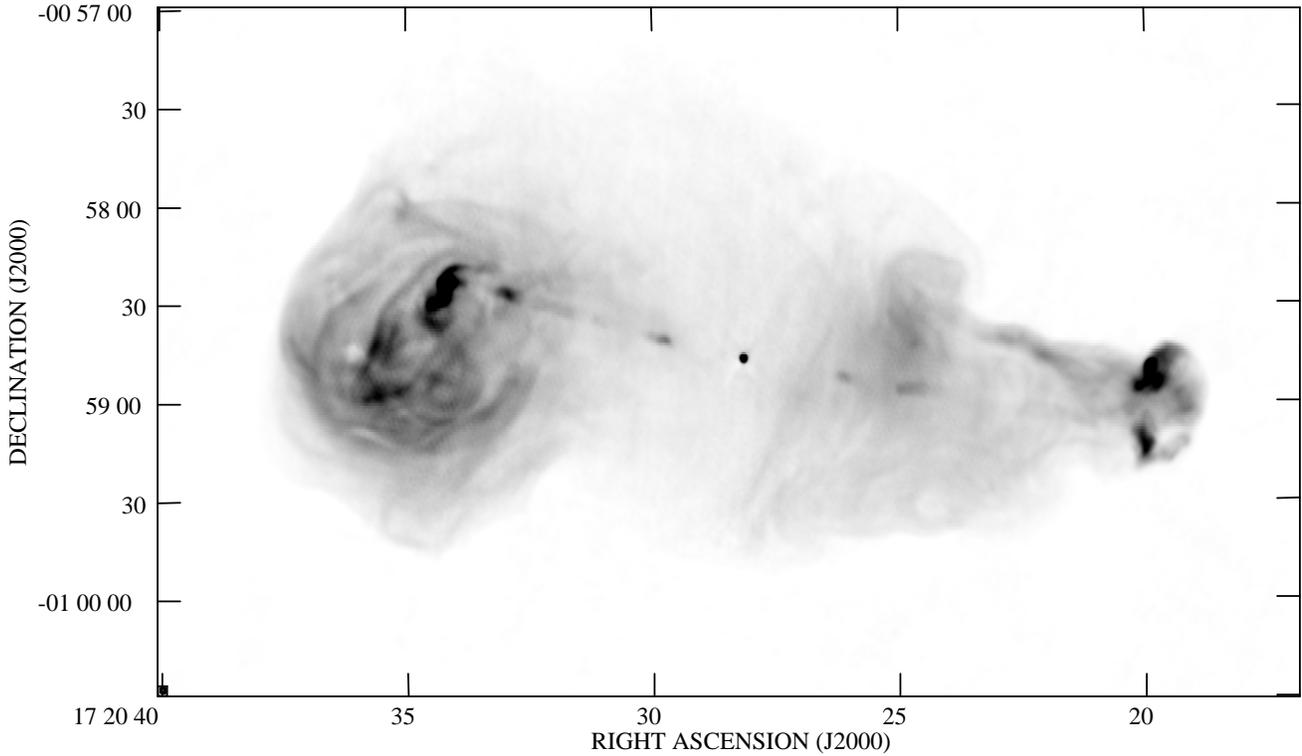}
\caption{1.67 GHz total intensity image of 3C\,353 at 1.8 arcsec resolution}
\label{fig:radiojlg}
\end{figure*}

\subsection{Radio}

We obtained VLA observations at 1.6\,GHz and 4.8\,GHz from the VLA
archive as well as carrying out new observations at 327\,MHz at the
VLA and at 620\,MHz at the NCRA Giant Meter-wave Radio Telescope
(GMRT) in 2006.  We also obtained images of 3C\,353 at 74\,MHz and
8.4\,GHz (see below).  All the radio data were reduced in
\textsc{aips}. The radio observations are summarized in
Table~\ref{tab:obsjlg} and Fig.~\ref{fig:radiojlg} shows an example of
the image quality achieved.

\begin{table*}
  \centering
    \caption{Radio Observation Details}
    \begin{tabular}{c|c|c|c|c|c|c|c|c}
      \hline
      Band & Frequency & Telescope:Config. & Date & Duration & Bandwidth & Phase Calibrator & Dynamic Range & Res. \\
       & (MHz) & & & (s) & (MHz) & & & (arcsec)\\ 
      \hline
      4 & 73.8 & VLA & 07/03/1998 & 7740 & 1.3 & 3C405 & 2090:1 & $25.0\times25.0$\\[3pt]
      P & 327.3 & VLA:A+PT & 27/05/2006 & 25660 & 3.125 & 3C286  & 580:1 & $6.75 \times 2.92$ \\
       & & VLA:B & 04/10/1998 & 3010 & 0.098 & 1416$+$067 & & \\[3pt]
      - & 614 & GMRT & 19/07/2006 & 19847 & 16.0 & 1743-038 & 590:1 & $6.41 \times 4.27$ \\[3pt]
      L & 1665/1385 & VLA:A & 19/05/1968 & 20710 & 12.5 & NRAO530 & 500:1 & $1.78 \times 1.44$\\
      & & VLA:B & 14/08/1968 & 6640 & 25.0 & NRAO530 & & \\
      & & VLA:C & 16/09/1985 & 1441 & 50.0 & 1741-038 & & \\[3pt]
      C & 4848/4898 & VLA:B & 14/08/1986 & 6799 & 50.0 & NRAO530 & 1180:1 & $2.24 \times 1.81 $ \\
      & & VLA:C & 22/08/1985 & 1679 & 50.0 &  1741-038 & & \\
      & & VLA:D & 15/11/1993 & 532 & 50.0 & 1730-130 & & \\
      & & VLA:D & 28/09/1984 & 3360 & 50.0 & 1725+044 & & \\[3pt]
      X & 8440/8452 & VLA:BCnD & 13/03/1994 & & 12.5 & & 2500:1 & $1.30 \times 1.30$\\
      \hline
    \end{tabular}
  \label{tab:obsjlg}
\end{table*}

\subsubsection{4 Band; 74\,MHz, 4\,m}

The 74-MHz observations used are those discussed by
\citet{kassim07jlg} and the data reduction is described in that paper.

\subsubsection{P Band; 327\,MHz, 91\,cm} 

Our 327\,MHz observations of 3C\,353 were taken at the VLA on the 27th
May 2006 using the nearby Pie Town antenna of the Very Large Baseline
Array (VLBA) to increase the long baseline of our A configuration data
set by a factor of two.  We carefully monitored the Pie Town antenna
during the calibration steps to ensure that the long baselines
survived.

Initial bandpass calibration using 3C\,286 resulted in smooth phase
and amplitude variation across all channels.  As a flux calibrator,
3C286 is slightly resolved at this frequency. If the fractional
bandwidth is great enough that the intrinsic visibility of 3C\,286
varies across the pass band, a model is required \citep{lazio05jlg}.
At 327\,MHz with the Pie Town antenna, we found that this effect is
negligible so 3C286 was used as the flux calibrator without a model,
effectively treated as a point source.  3C\,286 was used to calibrate
the phase across all baselines.  The resultant calibrated data set was
flagged to remove noise then spectrally averaged using \textsc{splat},
averaging every 4 channels to reduce the number of channels to 6,
whilst producing a single source data set. The calibrated data
included baselines up to 80\,k$\lambda$, twice what would have been
achieved with only the A configuration.

The B-configuration data set was bandpass calibrated using 3C\,286 to
calibrate the channel gains, then flux using 3C\,286 and phase
calibrated with $1416+067$.  After flagging and phase
cross-calibrating with the A configuration data set, the data sets
were merged and a multi-facet deep cleaned map of 106 fields was
produced.  The dynamic range of the final map is 580:1 with a
resolution of $6.7 \times 2.9$ arcsec.

\subsubsection{620\,MHz, 48\,cm}

Our 620\,MHz observations were taken at the GMRT on 19th July 2006.
The two observing frequencies (upper and lower sidebands) were
calibrated separately.  Preliminary bandpass calibration was performed
using 3C\,286 followed by flagging and one iteration of {\it flgit} to
remove RFI from all sources.  This flagged data set was then bandpass-
and flux-calibrated before being manually flagged and another
iteration of {\it flgit} applied.  The data set was spectrally
averaged into a multi-source data set, cutting the number of channels
from 128 to 6, whilst applying the bandpass calibration.  This data
set was finally flux- and phase-calibrated before 3C\,353 was {\it
split} into a single-source file.  The flux and phase calibrator used
was 1743-038.  One iteration of phase self-calibration was performed
before the data from the two observing frequencies were split into
individual channels and recombined using \textsc{dbcon} to a single
data set containing all of the averaged channels. This combined data
set was phase self-calibrated twice more and the final image has
resolution $7.50 \times 4.41$ and dynamic range $590:1$.

\subsubsection{L Band; 1.67\,GHz, 22\,cm}

Observations from A-, B- and C-configurations were flux- and
phase-calibrated separately using NRAO530 as the phase calibrator.
Due to the allocation of the observing frequencies in the
C-configuration data set, the data at the two frequencies were split
and recombined using \textsc{dbcon} to match the frequency allocations
of the A- and B-configuration data sets.  Once combined, the data were
phase self-calibrated and deep cleaned, with 100,000 iterations of
\textsc{clean} applied.  The final image has a resolution of 1.7
$\times$ 1.4 arcsec and a dynamic range of $500:1.$

\subsubsection{C Band; 4.8\,GHz, 6.3\,cm}

For maximum {\it uv} coverage, we use B-, C- and D-configuration data
sets, each of which were phase- and flux-calibrated and flagged.
We combine two epochs of D-configuration observations to increase
the signal to noise. 

The final combined data set was phase and amplitude self-calibrated,
normalizing the gain.  There was no significant reduction in the
maximum flux when compared to phase-only self-calibration, but the
image quality was improved.  The final image has a resolution $2.24
\times 1.81$ arcsec and a dynamic range of 1180:1.

\subsubsection{X Band; 8.4\,GHz, 3.6\,cm}

The X band images of 3C\,353 were kindly provided by Alan Bridle.  The
deep cleaned images have a dynamic range of ~2500:1 and a resolution
of $1.30 \times 1.30$ arcsec.  The details of the reduction are
described by \citet{swain96jlg}.

\begin{figure*}
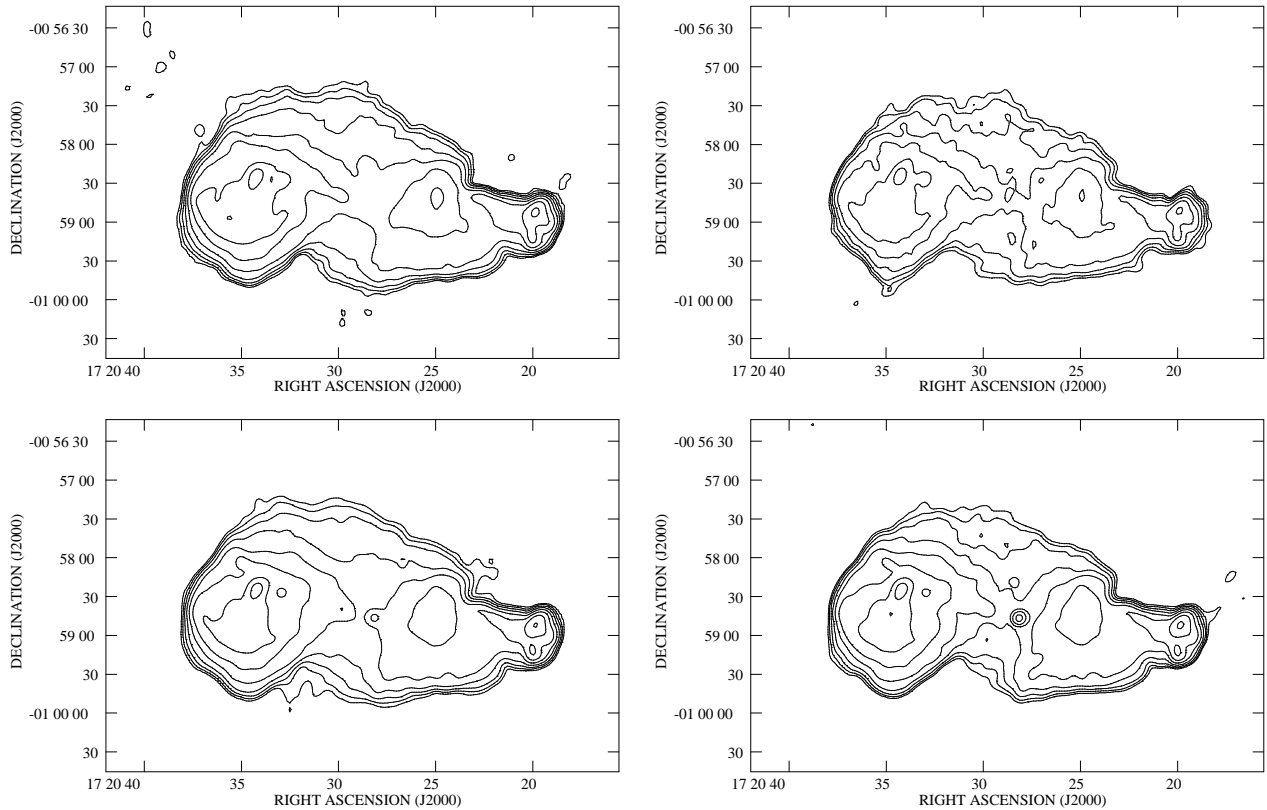

\includegraphics[width=0.48\textwidth]{3C353-Pcont5s-jlg.ps}
\includegraphics[width=0.48\textwidth]{3C353-6cont5s-jlg.ps}
\includegraphics[width=0.48\textwidth]{3C353-Lcont5s-jlg.ps}
\includegraphics[width=0.48\textwidth]{3C353-Ccont5s-jlg.ps}
\caption{Radio contour images of 3C\,353 at 327\,MHz (top left), 620\,MHz (top right), 1.67\,GHz (bottom left) and 4.8\,GHz (bottom right) all convolved to 7.5$\times$7.5 arcsec resolution.  The contour levels are at $5\sigma\times$(1,2,4,...)\,mJy/beam where $5\sigma = $ 0.0106 for 327\,MHz,  0.0113 for 620\,MHz, 0.0045 for 1.67\,GHz and 0.0017 for 4.8\.GHz }
\label{fig:contjlg}
\end{figure*}

\subsection{X-ray}

We observed 3C\,353 on the 25th August 2006 and 17th February 2007 with
{\it XMM-Newton} EPIC MOS1, MOS2 and pn cameras.  The initial
observation (Set 1) from 25th August 2006 yielded 39\,522\,s and
39\,525\,s for MOS1 and MOS2 respectively and 34\,042\,s for the pn
camera, whereas the second observation on the 17th February 2007 (Set
2) was for 10\,473\,s and 10\,487\,s for MOS1 and MOS2 and 5\,653\,s
for pn.  The pn camera was in Extended Full Frame mode for both
observations and the MOS cameras were in Full Frame mode.

The data sets were initially processed using \textsc{sas} version
7.0.0. Standard EPIC MOS and PN pipelines and the standard filters
\textsc{$\#$xmmea$\_$ea} and \textsc{$\#$xmmea$\_$ep} were applied.
The MOS data sets were filtered to include single, double, triple and
quadruple events (PATTERN$\le$12) whereas the PN data set was filtered
to include only single and double events (PATTERN$\le$4).  Set 1 MOS
data sets were free from flare events whilst the PN data set was
filtered with a count threshold of 2 counts per second to remove a
small flare event at the end of the observation.  In Set 2, all
cameras were affected by flaring; the MOS and PN data sets were
filtered with count rate thresholds of 0.3 and 0.7 counts per second
respectively.  As only 1510 seconds remained of the PN data, this data
set was excluded from further analysis.  The net livetimes for Sets 1
and 2 were 44\,039 for MOS1 and MOS2 and 26\,052 for PN.

The MOS and PN data sets were energy filtered to include energies
between 0.5 and 5\,keV before using the \textsc{sas} command {\it
evselect} to generate images from the filtered events files. The task
{\it eexpmap} was then used to generate exposure maps without
vignetting.  We chose not to include vignetting correction in these
images as it incorrectly weights up the particle background at the
edge of the field, creating artefacts in the image.  The images from
the separate detectors were normalized to the Set 1 PN count rate (in
order to remove the PN chip gaps) and then combined.  The exposure
maps were combined then divided by the total filtered livetime.  The
combined normalised image was finally divided by the scaled combined
exposure map.  The combined image was Gaussian smoothed with a kernel
of 10.4\,arcsec to highlight the clusters and radio lobes using the
\textsc{ciao} command {\it aconvolve}.

X-ray spectra were extracted using {\it especget}, a script that
combines the {\it evselect}, {\it arfgen} and {\it rmfgen} commands in
\textsc{sas}, from regions defined to examine the emission from the
northern and southern sub-cluster regions, and the east and west
lobes.  As the core and lobes are small regions near the pointing
centre, so that vignetting is not important, we used local background
subtraction.  However, for the cluster, vignetting needs to be
considered.  The filtered event files were weighted using {\it
evigweight} before spectra were extracted using a double subtraction
method, \citep[e.g.][]{arnaud02jlg}.  Background template files for
the field of view were made using files created by \citet{read03jlg}
and scaled to the same particle background level as our observations.
A local background region away from the galaxy and cluster was defined
using \textsc{ds9}.  A template background spectrum was subtracted
from the source spectrum events file to account for instrument and
particle noise before a local background region was used to subtract
residual background emission due to the differences in the
Galactic/extragalactic background levels of the source and scaled
background datasets.  Due to the low signal to noise of Set 2, only
Set 1 could be used in the double subtraction method applied to the
cluster.  The spectra were then binned to 20 counts per channel after
background subtraction, ignoring the first 20 channels for the MOS
cameras and the first 50 channels for the pn camera.

\begin{figure*}
\includegraphics[width=0.48\textwidth]{3C353-xrbw8-jlg.ps}
\includegraphics[width=0.50\textwidth]{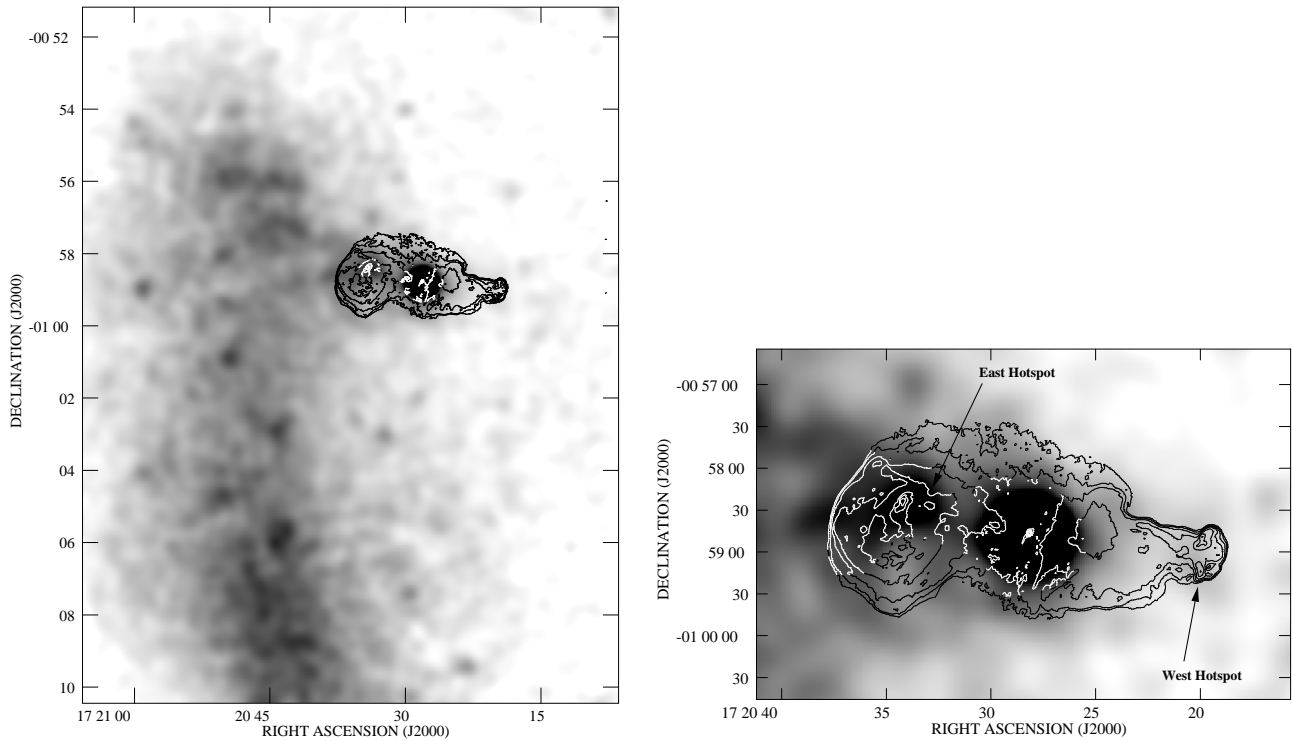}
\caption{Radio contours of 3C\,353 at 1.67 GHz with a 0.3-7.0 keV {\it XMM-Newton} image (MOS 1, MOS 2 + pn) of the X-ray emission from the radio lobes and the cluster Zw\,1718.1-0108 (left) Gaussian smoothed with a kernal of 10.4\,arcsec; the same image zoomed in to show the inverse-Compton emission from the lobes (right), {\edit with the East radio/X-ray hotspot and West radio hotspot labelled.}}
\label{fig:xrayjlg}
\end{figure*}

\section{Results}
\label{sec:resultsjlg}

In this section we discuss the results of the X-ray spectral fitting
and radio flux density measurements.  X-ray fitting was carried out
using {\sc xspec} v 11.3 in the energy range 0.3 -- 7.0 keV.  Where
thermal models were fitted we used a redshift of 0.03.

\subsection{Absorbing column density}
\label{sec:nhjlg}

The column density towards 3C\,353 is uncertain.  \citet{iwasawa00jlg}
adopt a value of $1.0 \times 10^{21}$\,cm$^{-2}$ based on the {\sc
H\,i} measurements of \citet{dickey90jlg}.  However, they note that
the visual extinction in the direction of 3C\,353 would correspond to
a column density of a factor $\sim 2$ higher.  From the Galactic dust
measurements of \citet{schlegel98jlg} we estimate an $A_{V}$ of 1.4
mag, which would correspond to a column density of $\sim 2.6 \times
10^{21}$ for standard Galactic gas/dust ratios.  The true value is
likely to lie somewhere between these extremes.

We therefore used our new data to estimate a column density directly.
We extracted spectra for bright point sources around 3C\,353 using
local background regions.  In Set 1 there were 5 point sources bright
enough for spectroscopy.  We fitted power-law models with free
absorption to these sources and found a weighted mean $N_{H}= (1.64 \pm
0.07) \times 10^{21}$cm$^{-2}$.  As this is consistent with the range
estimated above, we adopt the mean value in the analysis that follows.

The X-ray point sources are not shown in Fig.~\ref{fig:xrayjlg} and
were excluded from all spectra extracted where the cluster was
examined. The emission from the radio galaxy was also excluded.

\subsection{The Nucleus}
\label{sec:corejlg}

In the X-ray, the nuclear region was defined with a radius of 40 arcsec
to include as many nuclear photons as possible while excluding the X-ray
emission from the radio lobes.  The background was estimated from a
local background region positioned above the radio galaxy so as not to
include any emission related to the radio galaxy.  In the radio, the
nuclear region was much smaller, adjusted in each data set to be a close
fit to the nuclear emission.  The peak intensity was determined using
{\it jmfit} for each frequency, the results of which are shown in
Table~\ref{tab:radioSjlg}.

\begin{figure}
\includegraphics[width=0.36\textwidth,angle=270]{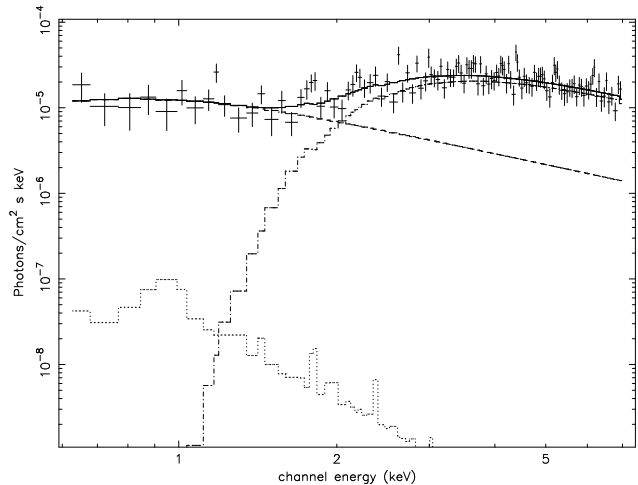}
\caption{X-ray spectrum of the nucleus in the energy range 0.3--7.0\,keV. The model plotted is the dual power-law + thermal model as described in Section~\ref{sec:corejlg}.  For clarity only the PN data are plotted.}
\label{fig:corespectrajlg}
\end{figure}

{\edit We initially attempted to fit single power-law and thermal
models to this spectrum; however, the fits were poor ($\chi^{2}=466.6$
for 259 d.o.f. and $\chi^{2}=2242.5$ for 259 d.o.f. respectively).  We
therefore fitted a model which has been shown \citep{hardcastle06jlg}
to provide a good fit to the nuclear X-ray emission from narrow-line
radio galaxies,} consisting of a power-law component at Galactic
absorption (see Section~\ref{sec:nhjlg}), a second power-law component
with redshifted intrinsic absorption and a $kT=1$\,keV thermal
component.  The best fitting model had $\chi^{2}=268$ for 237
d.o.f. {\edit The core spectrum is shown in
Fig.~\ref{fig:corespectrajlg}.}  The photon indices of the unabsorbed
and absorbed power-law components were $\Gamma = 1.49 \pm 0.1$ and
$1.33 \pm 0.3$ respectively, and the intrinsic column density was $6.1
\pm 0.6 \times 10^{22}$\,cm$^{-2}$.  The unabsorbed X-ray component
has a 1\,keV luminosity density of $1.86 \times
10^{16}$\,W\,Hz$^{-1}$\,sr$^{-1}$, whereas the absorbed X-ray
component with the absorption removed and extrapolated to include
energies from 2--10\,keV has a luminosity of $2.82 \times
10^{42}$\,erg\,s$^{-1}$.  We discuss the interpretation of the nuclear
emission in Section~\ref{sec:dis-corejlg}.

\begin{table}
  \centering
    \caption{Radio Core Fluxes}
    \begin{tabular}{c|c|c|c}
      \hline
      Band & Frequency & Flux & Luminosity\\
      & (MHz) & (Jy) & (W\,Hz$^{-1}$\,sr$^{-1}$)\\
      \hline
      P & 327.3/329.6 & 0.097 & 1.5$\times 10^{22}$\\
      - & 620.5/633.3 & 0.089 & 1.4$\times 10^{22}$\\
      L & 1665/1678 & 0.116 & 1.8$\times 10^{22}$\\
      C & 4848/4898 & 0.140 & 2.2$\times 10^{22}$\\
      X & 8440/8452 & 0.149 & 2.3$\times 10^{22}$\\
      \hline
    \end{tabular}
  \label{tab:radioSjlg}
\end{table}

\subsection{Lobe properties}
\label{sec:lobesjlg}

By considering the X-ray emission from regions encompassing the east
and west lobes but excluding the hotspots and the core, we extracted
spectra from the combined X-ray data set using local background
regions which we fitted with both thermal emission models and power
laws.  Assuming a circularly symmetric model for the core, we
determined that $\sim 6$ per cent of the counts in the East lobe
spectra are from the core whereas the core counts account for a
quarter of those in the West lobe spectra.  Using the {\it XMM-Newton}
calibration files, we determined the fraction of the core spectrun
expected to be scattered into the West lobe region to be 4 per cent of
the total core emission, and so we included a fixed component in the
West lobe spectral fit consisting of the best-fitting core model with
a normalisation fixed at 0.04 times that of the core.  The details of
the core spectrum are given in Section~\ref{sec:corejlg}.  Due to the
low signal to noise of Set 2, the West lobe was undetected in this
data set and so our spectra were extracted from Set 1 only.  

{\edit The background emission from the cluster environment is the
primary contributor to the uncertainty in the lobe flux densities.  To
best take account of this variation we used a truncated annular
background region for the East lobe where the lobe appears to interact
with the cluster.  We investigated the effect of using local
background regions above and below the galaxy; the best fitting
power-law models gave consistent photon indices but led to an increase
of a factor 2 in the flux density of the East lobe. These background
regions only take account of the emission from one of the two
sub-clusters leading us to believe that a truncated annular background
region is a more accurate measure of the cluster's contribution.  For
the West lobe, a local background region located to the south of the
galaxy, at a similar distance from the brightest cluster emission took
account of the emission from the southern sub-cluster sufficiently to
allow a good fit to the spectra.}

\begin{figure*}
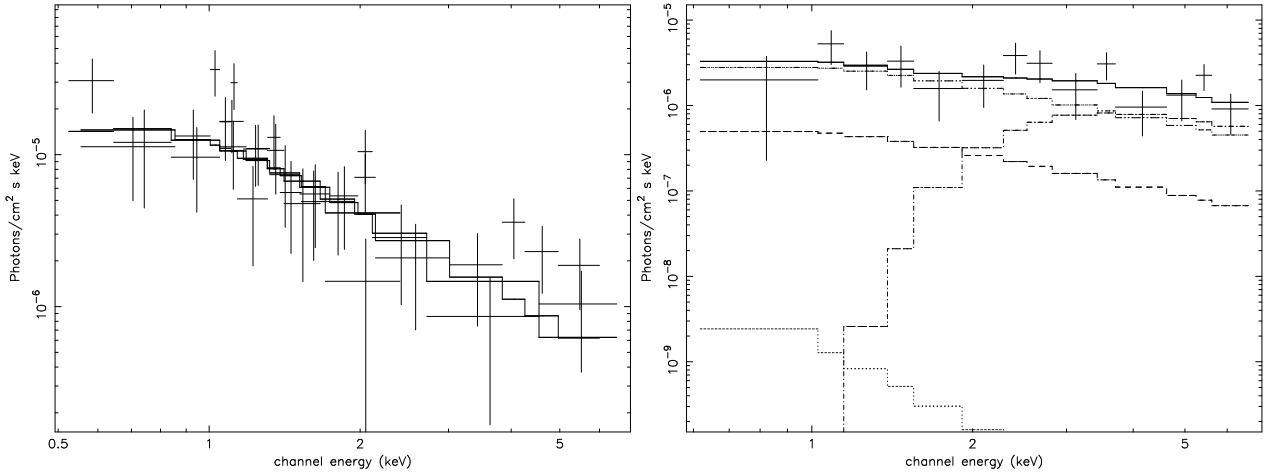

\includegraphics[width=0.35\textwidth, angle=270]{3C353E-spectra-jlg.ps}
\includegraphics[width=0.35\textwidth, angle=270]{3C353W-spectra-jlg.ps}
\caption{X-ray spectra for the East (left) and West (right) lobes in the energy range 0.3--7.0\,keV.  The East lobe shows the MOS 1, MOS 2 and PN cameras with the power-law model while the West lobe shows only the PN data for clarity with the dual power-law + thermal contribution from the core as well as the best-fitting power-law model from the lobe emission as described in Section~\ref{sec:lobesjlg}.}
\label{fig:spectrajlg}
\end{figure*}

The power-law model gave better fits than the thermal model for both
lobes with $\chi^{2}=39.6$ for 34 d.o.f for the East lobe and
$\chi^{2}=23.1$ for 24 d.o.f for the west lobe.  The photon indices
were $\Gamma = 1.9 \pm 0.4$ and $\Gamma = 1.2 \pm 0.6$ respectively,
coresponding to $\alpha=0.9 \pm 0.4$ and $\alpha=0.2 \pm 0.6$. {\edit
Figure~\ref{fig:spectrajlg} shows the best-fitting power-law model for
the East and West lobe X-ray data.}  The low frequency X-ray spectral index,
measured between 327\,MHz and 1.67\,GHz, is $\sim 0.7$ for both lobes,
so that the X-ray spectral indices of both lobes are consistent with a
CMB inverse-Compton scattering model.  We discuss the nature of the
lobes in Section~\ref{sec:dis-lobesjlg}.

\begin{figure*}
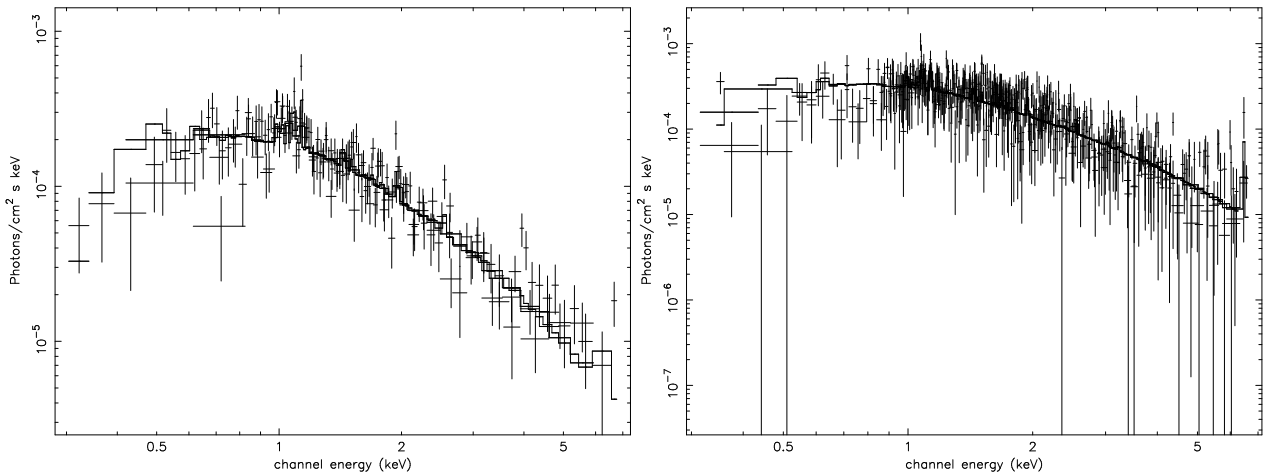

\includegraphics[width=0.35\textwidth,angle=270]{3C353N-spectra-jlg.ps}
\includegraphics[width=0.35\textwidth,angle=270]{3C353S-spectra-jlg.ps}
\caption{X-ray spectra for the northern (left) and southern (right) sub-clusters in the energy range 0.3--7.0\,keV.  The MOS1, MOS2 and PN data sets are shown in each case with the best-fitting thermal model as described in Section~\ref{sec:clusterjlg}. }
\label{fig:clusterspectrajlg}
\end{figure*}

\subsection{The nature of the hotspots}
\label{sec:hotspotjlg}

Both the East and West hotspots are clearly detected in all radio
frequencies.  The West hotspot is undetected in the X-ray; however, the
East hotspot has an X-ray counterpart $6.5 \pm 0.8$ arcsec ($4.0 \pm
0.5$\,kpc) from the centre of the radio hotspot.  This has too low
signal to noise for spectral analysis to be performed, with only 60
counts between 0.3 and 7.0\,keV.  In recent {\it Chandra} observations
(Kataoka et al., private communication), the same feature is detected
but the offset was measured as $4.6 \pm 0.5$ arcsec.  A jet knot was
also detected in the vicinity of the East X-ray hotspot suggesting
that the difference in our offset measurements could be due to
contamination from this knot.

The inverse-Compton analysis performed on the lobes cannot be applied
here as the electron populations responsible for any inverse-Compton
X-ray and radio synchrotron emission reside in different regions.
X-ray hotspots in radio galaxies of similar radio luminosity to
3C\,353 are generally thought to be produced by synchrotron emission,
and so this is likely to be the dominant emission process for this
X-ray hotspot as well.  The offsets measured between the X-ray and
radio hotspots are consistent with what is seen in some other sources
\citep{hardcastle02jlg,hardcastle07jlg,erlund07jlg}.

\subsection{Cluster characterization}
\label{sec:clusterjlg}

Unfortunately, the southernmost section of the southern sub-cluster
extends beyond the field of view of the {\it XMM-Newton} cameras;
however, a comparison to the {\it ASCA} image of \citet{iwasawa00jlg}
shows that the peak of the X-ray surface brightness is included in our
data.

\begin{figure}
\includegraphics[width=0.47\textwidth]{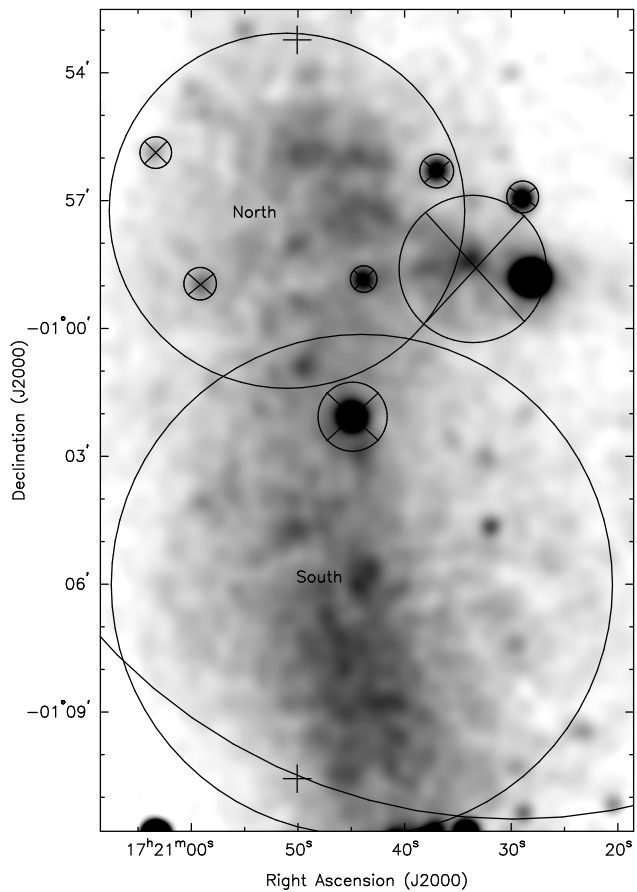}
\caption{Northern and southern sub-cluster regions from which spectra were extracted to determine the sub-cluster global temperatures.  The crossed regions are those which were excluded from the analysis and include the X-ray point sources and 3C\,353.  The crosses indicate the limits of the slice used.  The inner edge of a large annular region masking the edge of the camera can be seen at the bottom of the image.}
\label{fig:regionsjlg}
\end{figure}

\begin{figure}
\includegraphics[width=0.34\textwidth,angle=270]{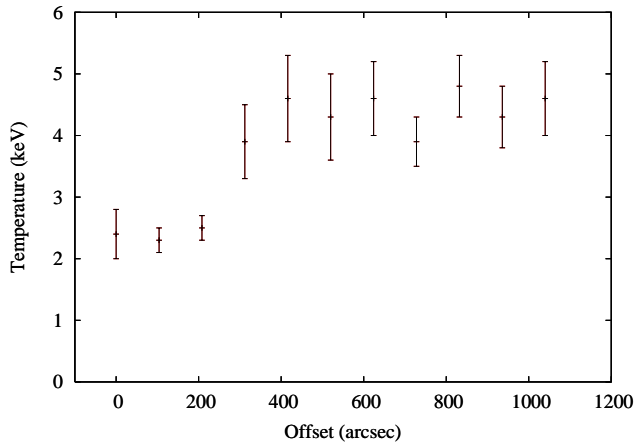}
\caption{Temperature variation of north-south slice through cluster with rectangular regions of  $208 \times 104$\,arcsec from a zero of offset at 17:20:50.074, -00:53:13.77.}
\label{fig:tempprofjlg}
\end{figure}

We extracted spectra from the northern and southern sub-clusters using
a double subtraction method with annular local background regions.
The northern and southern sub-cluster regions are shown in
Fig.~\ref{fig:regionsjlg}.  Thermal models fitted using \emph{mekal}
in \textsc{Xspec} give {\edit acceptable} fits ($\chi^{2} = 258$ for
175 d.o.f and $\chi^{2} = 544$ for 490 d.o.f for the northern and
southern sub-clusters respectively) with $kT= 3.3 \pm 0.3$\,keV for
the northern sub-cluster and $kT= 4.0 \pm 0.5$\,keV for the southern
sub-cluster both fitted with an absorption of $N_{H} = 1.6 \times
10^{21}$\,cm$^{-2}$.  {\edit The best-fitting models are shown in
Fig.~\ref{fig:clusterspectrajlg}.}  \citet{iwasawa00jlg} determined
the temperature of the combined sub-clusters to be $kT = 4.3 \pm
0.2$\,keV using a Galactic absorption of $N_{H} = 1.0 \times 10^{21}$
cm$^{-2}$ whereas when the upper limit ($1.6 \times 10^{21}$cm$^{-2}$)
was used, the temperature dropped to $3.94 \pm 0.21$\,keV which is
consistent with the mean temperature of the cluster.  The difference
in temperature of $\sim 1$\,keV supports the idea that the two
sub-clusters are originally separate components undergoing a merger.
Taking a north-south slice through the cluster using rectangular
regions of $208 \times 104$\,arcsec from 17:20:50.074, -00:53:13.77
(Fig.~\ref{fig:tempprofjlg}), there is no {\edit evidence for a shock
and no increase in surface brightness, indicating a non-violent
}interaction between these components.

A thermal model was fitted to annular regions in the northern and
southern sub-clusters.  The results are shown in
Table~\ref{tab:tempsjlg}.  The residual local background was accounted
for using regions in a relatively source free area of the field of
view.  Using an annular background region gave consistant temperatures
for both sub-clusters. The northern sub-cluster shows a linear
increase in temperature with radius with a best fit of $T(r)=0.010r +
1.528$\,keV while the southern sub-cluster is isothermal within the
errors.  3C\,353's East lobe lies at a radius of 250 -- 350 arcsec
from the centre of the northern sub-cluster but it may also be affected by
the southern sub-cluster.  The annular temperature at this radius ($kT
= 3.5 \pm 0.5$\,keV) is consistant with the global northern
sub-cluster temperature ($kT = 3.3 \pm 0.3$\,keV) as well as the
temperature measured by the slice at this radius ($kT = 3.9 \pm
0.6$\,keV).  We therefore adopt a temperature of $3.5 \pm 0.5$\,keV
for the environment of 3C\,353.

\begin{table}
  \centering
    \caption{Temperature profiles for the northern and southern sub-clusters}
    \begin{tabular}{c|c|c||c|c|c}
      \hline
      \multicolumn{3}{|c|}{Northern} & \multicolumn{3}{|c|}{Southern} \\
      Radius & $kT$ & $\chi^{2}$/d.o.f. & Radius & $kT$ & $\chi^{2}$/d.o.f.\\
      (arcsec) & (keV) & & (arcsec) & (keV) & \\
      \hline
      0 -- 50 & $1.8 \pm 0.3 $ & 29/19 & 0 -- 75 & $5.2 \pm 0.7 $ & 537/348\\
      50 -- 100 & $2.2 \pm 0.3 $ & 121/65 & 75 -- 150 & $4.2 \pm 0.3$ & 638/444\\
      100 -- 150 & $2.8 \pm 0.4 $ & 130/102 & 150 -- 200 & $4.7 \pm 0.4 $ & 250/167\\
      150 -- 250 & $3.5 \pm 0.3 $ & 129/108 & 200 -- 250 & $4.9 \pm 0.5 $ & 296/183\\
      250 -- 350 & $3.5 \pm 0.5 $ & 286/172 & & & \\
      \hline
    \end{tabular}
  \label{tab:tempsjlg}
\end{table}

\begin{figure}
\includegraphics[width=0.48\textwidth]{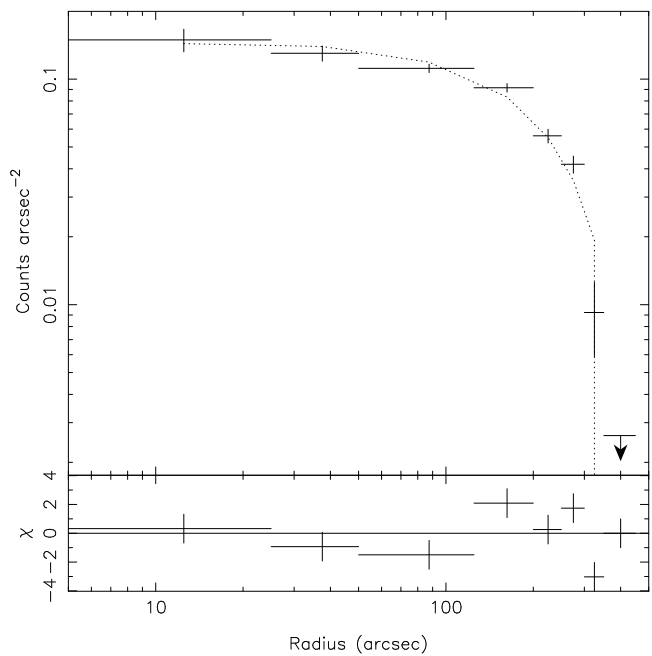}
\caption{Surface brightness profile for the northern sub-cluster showing the PN data set with the best-fitting $\beta$ model which has $\beta = 0.35$ and core radius 194 arcsec.}
\label{fig:radialjlg}
\end{figure}

As 3C\,353 interacts mainly with the northern sub-cluster and as the
southern sub-cluster is not fully covered within the {\it XMM-Newton}
field of view, surface brightness and pressure profiles were
constructed for the northern sub-cluster only.  To determine the
centre of the northern sub-cluster we used a centroiding routine for
each of the {\it XMM-Newton} cameras. Annular regions were used to
extract a surface brightness profile, excluding the chip gaps, X-ray
point sources, emission from 3C\,353 and from the southern sub-cluster
and the position of the missing chip in the MOS 1 camera.  An annular
region was used to account for the residual local background emission.
The surface brightness profiles for each camera were fitted with a model
consisting of a $\beta$ model convolved with the {\it XMM-Newton} PSF based
on the on-axis parametrisation described in the {\it XMM-Newton} CCF files
XRT1\_XPSF\_0006.CCF, XRT2\_XPSF\_0007.CCF and XRT3\_XPSF\_0007.CCF
before being fitted with a $\beta$ model (Fig.~\ref{fig:radialjlg}).

We examined the annuli in quadrants to highlight any density variation
in a north-south or east-west direction.  The northern quadrants and
the south-east quadrant were consistent.  We excluded the south-west
quadrant from the final profile to ensure that X-ray emission from the
radio galaxy, and any local disturbances in the cluster gas, were
removed.  The resulting surface brightness profiles were consistent
for all cameras.  The joint best-fitting model, for all cameras, had a
radius of 194 arcsec and $\beta = 0.35$.  Considering a $1\sigma$
confidence interval for 2 interesting parameters, $\beta$ was
unconstrained (though we consider values below 0.35 and above 0.9
unrealistic), whereas the core radius of the northern sub-cluster was
limited to between 184 and 412 arcsec.  We discuss the results of this
section in Section~\ref{sec:dis-cluster}.

\section{Discussion}

\subsection{The nature of the core}
\label{sec:dis-corejlg}

The double-peaked nuclear X-ray spectrum, consisting of a heavily
absorbed component together with one with only Galactic absorption, is
typical of what is observed in narrow-line radio galaxies (NLRG)
\citep*[e.g.][]{sambruna99jlg,grandi06jlg,belsole06jlg,hardcastle06jlg}.
However, when we plot the unabsorbed luminosity of the heavily
absorbed nuclear component in 3C\,353 against its 178-MHz total radio
luminosity, as done by \citet{hardcastle06jlg} (their figure 3) we
find that 3C\,353 lies almost 1 order of magnitude below the
correlation they determined for NLRGs: that is, 3C\,353 is
overluminous in the radio for its X-ray luminosity or underluminous in
the X-ray for its radio luminosity, and lies closer to the region of
the plot occupied by low-excitation radio galaxies (LERGs). By
contrast, the unabsorbed component is consistent with the correlation
found by Hardcastle et al. between unabsorbed X-ray luminosity and
nuclear 5-GHz radio emission {\edit (their figure 3)}.

The optical emission-line classification of 3C\,353 is uncertain.
\citet{laing94jlg} use the ratio of the flux in the [\textsc{O\,iii}]
line to that in the H$\alpha$ to distinguish between the NLRGs and
LERGs.  The best published optical spectrum of 3C\,353 is that of
\citet{simpson96jlg}, and their quoted line fluxes clearly place
3C\,353 below the [\textsc{O\,iii}]/H$\alpha$ cutoff of 0.2 proposed
by Laing et al: thus by this definition 3C\,353 would be a LERG.
However, if the line fluxes are corrected for reddening using values
determined from the \citet*{schlegel98jlg} dust maps (see
Section~\ref{sec:nhjlg}), then the emission-line ratio rises to $0.17
\pm 0.02$, where the errors are derived from the normal 10 per cent
errors of Simpson et al. and are almost certainly too low: within the
errors we cannot say whether 3C\,353 should be classed as a LERG or
NLRG.  The nature of 3C\,353's nucleus both in the optical and in the
X-ray remains ambiguous.

\begin{figure*}
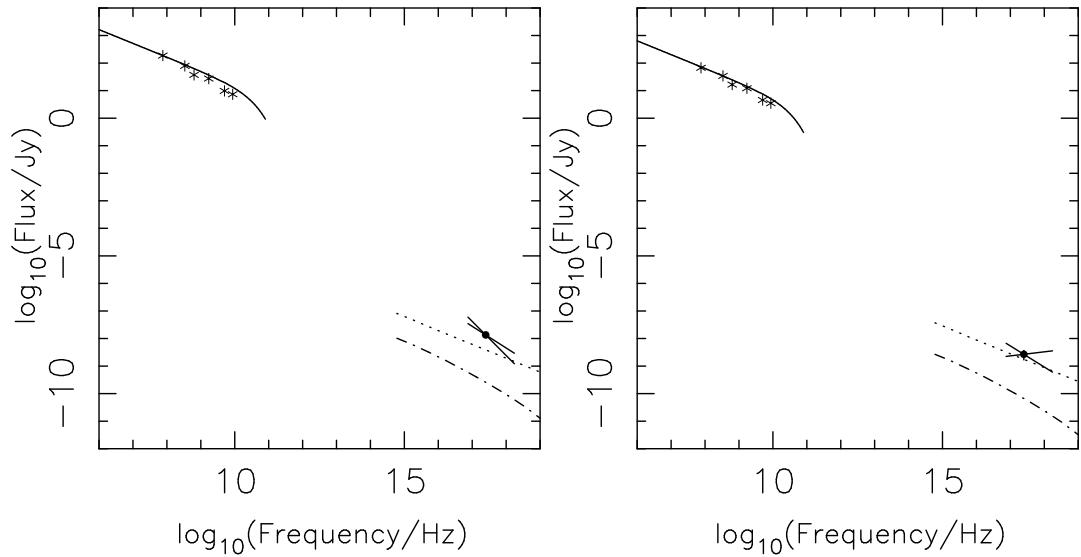

\includegraphics[width=0.40\textwidth]{3C353E-synch7-jlg.ps}
\includegraphics[width=0.40\textwidth]{3C353W-synch7-jlg.ps}
\caption{Broad-band spectrum for the East (left) and West (right) lobes of 3C\,353 . A synchrotron emission model (solid line) is fitted to the radio data (asterisks). The measured X-ray flux is represented by the dot with the bowtie indicating the error.  The dotted line shows the inverse-Compton X-ray flux prediction and the dash-dotted line shows the predicted synchrotron self-Compton emission, both with the equipartition magnetic field strength.}
\label{fig:synchjlg}
\end{figure*}

\subsection{Lobe emission and properties}
\label{sec:dis-lobesjlg}

Using the method of \citet{croston05jlg}, the X-ray observations were
compared to the predictions of an inverse-Compton model, based on
synchrotron modelling of the radio data sets using \textsc{synch}
\citep{hardcastle98jlg}.  The broad-band spectra for the lobes are
shown in Fig.~\ref{fig:synchjlg}.  The measured 1-keV flux densities
of the East and West lobes are $11.6 \pm 1.6 $ and $2.7 \pm 1.0$\,nJy.
The equipartition magnetic field strengths for the East and West lobes
are 0.89\,nT and 0.84\,nT respectively, while if we assume that all
the X-ray emission is inverse-Compton in origin then the measured flux
densities correspond to magnetic field strengths of $0.44 \pm
0.06$\,nT and $0.68 \pm 0.2$\,nT.  Therefore, while the weakly
detected West lobe has a measured magnetic field strength within
errors of the predicted equipartition value, the East lobe's measured
magnetic field strength exceeds the predicted equipartition value by a
factor $\sim 2$ so that $B_{obs}/B_{eq} \simeq 0.4$. The West lobe's
measured magnetic field strength is also consistent with a substantial
departure from equipartition.  This implies that the lobes of 3C\,353
are electron dominated, which is not unusual considering the range of
field strengths in the sample of \citet{croston05jlg}.  This factor
would be reduced for 3C\,353 if the lobes were not in the plane of the
sky.

The measured magnetic field strengths in the lobe correspond to
internal pressures of $3.4 \pm 0.6 \times 10^{-13}$\,Pa and $1.9 \pm
0.2 \times 10^{-13}$\,Pa for the East and West lobes respectively.
{\edit (If the alternative local background regions discussed in
Section~\ref{sec:lobesjlg} were used, the magnetic field strengths
would be reduced by $\sim 0.1$\,nT and the internal pressure of the
East lobe would double, but this does not affect our conclusions.)}

\begin{figure}
\includegraphics[width=0.48\textwidth]{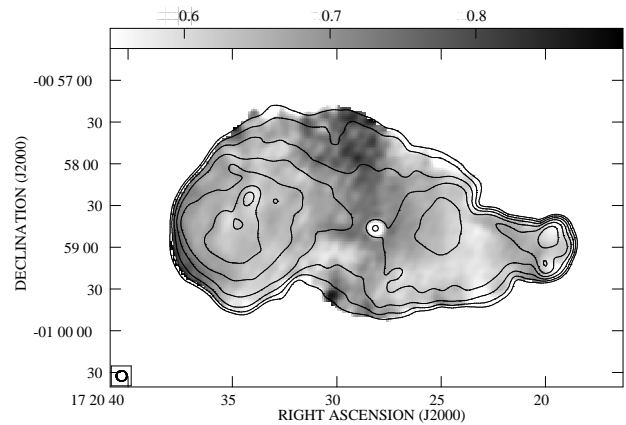}
\caption{Spectral index map between 1.67\,GHz and 327\,MHz with radio contours at 327\,MHz for levels of $0.01\times$(1,2,4,...)\,mJy/beam.  The beam size is shown by the circle in the bottom left corner and the colour bar at the top shows the mapping of the grey levels to spectral index.}
\label{fig:spixjlg}
\end{figure}

Using the radio data sets, we constructed a spectral index map between
the 1.67-GHz and 327-MHz images, Fig~\ref{fig:spixjlg}.  It revealed
a variation of $\Delta\alpha \sim 0.2$ across both lobes.  Within
the radio luminous region of the East lobe, excluding the hotspots,
the spectral index is roughly constant, $\alpha = 0.66 \pm 0.01$ despite the
filamentary structure seen in Fig.~\ref{fig:radiojlg}, whereas in the
West lobe, $\alpha = 0.63 \pm 0.01$ excluding the hotspot.  We note that
the radio lobes do not appear to be entirely separate, and thus
consider the region in between the radio luminous lobes, north and
south of the core to be an inter-lobe region, which cannot be
unambiguously associated with either lobe.  This inter-lobe region has
a relatively steep spectrum of $\alpha = 0.75 \pm 0.01$ whereas the hotspot
in both lobes exhibits a flatter spectrum of $\alpha = 0.60 \pm 0.01$.
Fig.~\ref{fig:spixjlg} shows the spectral index map between 1.67\,GHz
and 327\,MHz with contours from the 327-MHz map.  

The X-ray/radio ratio was determined for the hotspot, lobe and
inter-lobe regions, using the high signal to noise of the X-ray data
set.  The 327\,MHz radio map of resolution $7.0 \times 7.0$ arcsec was
used with the convolved, exposure corrected X-ray map of energies 0.3
to 7.0\,keV.  The X-ray/radio ratio was found to be a factor of 4
greater in the steep inter-lobe region than in the flat, hotspot
regions. If the magnetic field strength and the number densities of
both the inverse-Compton and synchrotron emitting electron are
constant across the lobe, the X-ray/radio ratio would also be
constant, but this is not what we see in 3C\,353. {\edit Similar
results have been seen in other radio galaxies, notably 3C\,452
\citep{isobe02jlg} and Pictor A \citep{hardcastle05jlg}.} Hardcastle
\& Croston consider the following three models to explain the varying
X-ray/radio ratio:
\begin{enumerate}
\item Some other emission process could boost the X-ray emission in
the inner regions;
\item The central regions contain more low-energy electrons relative
to the outer regions;
\item The magnetic field strength varies as a function of position.
\end{enumerate}

A contribution from the core was included in the X-ray spectral
analysis which also takes account of the contribution of the
unabsorbed X-ray emission associated with the jet and the absorbed
X-ray emission from the accretion disk.  Emission from the core cannot
therefore boost the X-ray emission in the inner regions. Our spectral
fits show no evidence for a contamination of the lobe spectra by
galaxy-scale thermal emission.  Neither could a boost be due to the
galaxy's proximity to the cluster as local background regions were
used to account for any thermal emission from the sub-clusters.  For
inverse-Compton scattering of nuclear photons to be the dominant
process, the core bolometric luminosity needs to be at least
$10^{38}$\,W.  This is not unrealistic, however modelling the lobe
surface brightness for this scenario using the results of
\citet{brunetti00jlg} as described by \citet{hardcastle02jlg} and with
3C\,353 in the plane of the sky, reveals a gradient across the lobe
which is not observed in either the X-ray image nor the X-ray/radio
ratio.  Using the spectral energy distributions of \citet{haas04jlg},
we deduced a typical IR spectrum for 3C\,353 by scaling a dual
power-law fit to the spectral energy distribution of 3C\,33.  3C\,33
was chosen as it has a similar luminosity and is classified as a
NLRG/FR\,II.  The IR spectrum was normalised by the ratio of
low-frequency radio luminosities of the two sources.  The predicted
flux density for the nuclear inverse-Compton emission is $6.5 \times
10^{-12}$\,Jy at 1\,keV, which is a factor of $\sim 700$ fainter than
the predicted CMB inverse-Compton emission and the observed flux
density ($4.4$ and $11.6$\,nJy respectively).  Thus we can rule out
model (i).

If we consider the magnetic field to be constant at the measured lobe
averaged magnetic field strength of 0.39\,nT and apply an
inverse-Compton model to the lobes, we find the emission at 327\,MHz
traces electrons with $\gamma \simeq 4000$ whilst the measured
inverse-Compton emission traces electrons at $\gamma \simeq 1000$.  As
the critical frequency for synchrotron emission goes as $\gamma^{2}$,
we find that a variation in the X-ray/radio ratio of a factor of 4
requires a variation in the spectral index between 10\,MHz and
327\,MHz of $\sim 0.5$.  Even if the equipartition magnetic field
strength of 0.89\,nT is assumed so that the 327\,MHz emission traces
electrons of $\gamma \simeq 3000$, the observed spectral index
variation requires a factor $\sim 1.2$ variation in the X-ray/radio
ratio which is still much lower than observed.  This is also the case
with Pictor A \citep{hardcastle05jlg}; a variation in the low-energy
electron densities alone (model ii) cannot explain the variation of
X-ray/radio ratio across the lobe.

\begin{figure}
\includegraphics[width=0.47\textwidth]{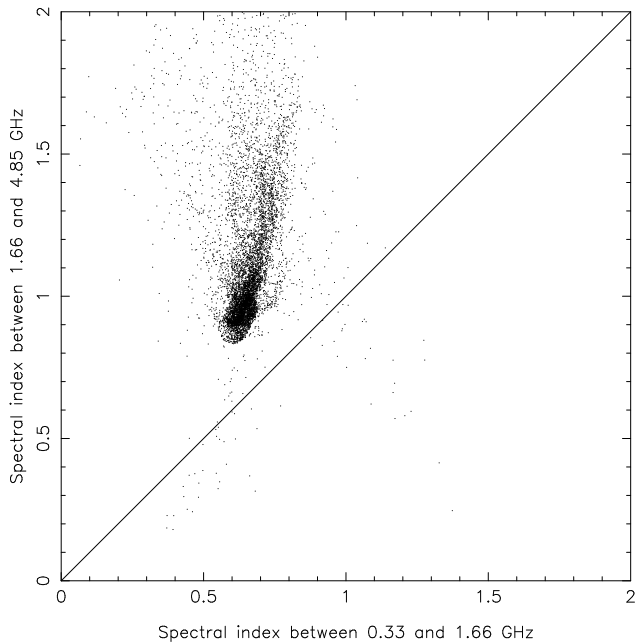}
\caption{Colour-colour diagram for 3C\,353 made at 7\,arcsec resolution.}
\label{fig:c-cjlg}
\end{figure}

Alternatively, if we assume constant electron densities for both the
synchrotron and inverse-Compton emission electrons and also that the
magnetic field strength does not vary along the line of sight, we find
that the observed variation in the X-ray/radio ratio then requires a
variation in the magnetic field of at most a factor of $~2.5$.  For a
given frequency, this means the spectral index observed at low
frequencies, between 1.67\,GHz and 327\,MHz, for regions of high
X-ray/radio ratio should correspond to the spectral index between
3.5\,GHz and 825\,MHz for low X-ray/radio ratio regions.  We consider
the spectral indices between 4.8\,GHz and 1.67\,GHz to limit the
spectral indices of the low X-ray/radio ratio regions and find that
they exceed the upper limit predicted by the observed X-ray/radio
ratio. From this comparison, we cannot rule out the possibility that a
varying magnetic field alone could be responsible for the observed
variations in the X-ray/radio ratio and the spectral indices.  We
therefore considered a colour-colour diagram using the method of
\citet{katz93jlg}.  In their study of Cygnus A, they argue that a
single curve on a colour-colour diagram is consistent with a
homogeneous distribution of relativistic electrons together with
varying the magnetic field strength across the source.  The
colour-colour diagram for 3C\,353, shown in Fig.~\ref{fig:c-cjlg}, has
a similar single curve and so is consistent with this picture.  In
Pictor A, \citet{hardcastle05jlg} argued that the detailed positional
variations of radio spectra and radio/X-ray ratio were not consistent
with such a model, but our data are not good enough to rule it out
here.

We conclude that a varying electron spectrum alone cannot account for the
observed variation in X-ray/radio ratio in the lobes of 3C\,353 but
that a magnetic field strength that varies by a factor of $\sim 2.5$
throughout the lobes can explain it.  Similar conclusions were reached
by \citet{hardcastle05jlg} in their study of Pictor A.

\subsection{The cluster-lobe interaction}
\label{sec:dis-cluster}

\begin{figure}
\includegraphics[width=0.47\textwidth]{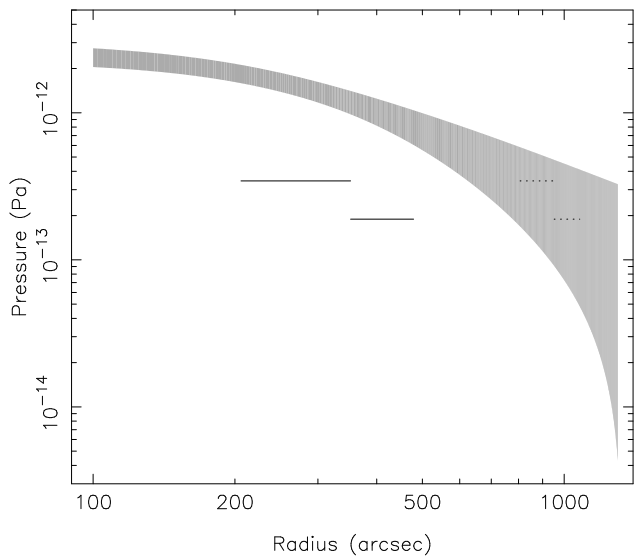}
\caption{Pressure profile for the northern sub-cluster shown by the greyed in region, with the measured pressure of the lobes of 3C\,353 at the projected distance (solid line) and at the position with the radial shift applied to where the East lobe is in pressure balance with the sub-cluster (dashed line).  The measured inverse-Compton emission was used to determine the pressure within each lobe.  The temperature of the cluster at the radial distance of 3C\,353 was used.}
\label{fig:pressjlg}
\end{figure}

The surface brightness profile was converted to a pressure profile
using the method of \citet{birkinshaw93jlg}, so that the radio lobe
pressures determined with \textsc{synch} could be directly compared to
the external pressure from the cluster at the position of the radio
galaxy.  The centre of the East lobe lies $\sim 280$\,arcsec from the
centre of the northern sub-cluster.  At this radial distance where the
temperature of the environment is taken as $3.5 \pm 0.5$\,keV, the
external pressure is much greater than the internal pressure of the
lobes {\edit (See Fig.~\ref{fig:pressjlg})}.  As FR\,II radio galaxies
are expected to be in pressure balance or over pressured, either
3C\,353 is not in the plane of the cluster or there is an additional
contribution to the pressure from non-radiating particles such as hot
thermal or relativistic protons in the lobes.  Previous studies of
FR\,II radio galaxies in groups and clusters have shown that pressure
balance can usually be achieved without additional protons
\citep[e.g.][]{croston05jlg, hardcastle02jlg, belsole04jlg}. We
therefore assume that 3C\,353 is not in the plane of the cluster and
using the East lobe, determine that 3C\,353 requires a shift in radial
position corresponding to $\sim 600$\,arcsecs to be in pressure
balance with the northern sub-cluster {\edit (here we use the best
estimate of the lobe pressure from Section~\ref{sec:dis-lobesjlg})}.
This places 3C\,353 $\sim 370$\,kpc either in front or behind the
centre of the cluster.  At this position, the West lobe's internal
pressure is also consistent with the intracluster pressure.  The
position of 3C\,353 with respect to the pressure profile is shown in
Fig.~\ref{fig:pressjlg}.  At this distance from the cluster, the
external pressures seen by the lobes are different by a factor of
$\sim 1.8$, which may help to explain the difference in appearance of
the lobes, the East having a spherical appearance whilst the West lobe
is elongated with hotspots at the outer edge.

\subsection{Additional features of note}

In addition to the usual radio-lobe structure, the East lobe of
3C\,353 contains a dark circular region of unknown origin at
17:20:36.088, -00:58:44.15 (Fig.~\ref{fig:radiojlg}).  Detectable in
all radio frequencies except 74\,MHz (presumably due to its low
resolution), the measured fluxes in this region are a factor of 100
above the rms background but are a factor of 2 fainter than the
surrounding lobe emission.  The X-ray image shows no sign of any
feature in this region and neither does the Digital Sky Survey
(DSS). As the level of the deficit is independent of frequency we can
rule out an foreground absorber and the required geometry is
unrealistic for an obstruction in the lobe.  Without further
information we are unable to identify this feature.

\citet{iwasawa00jlg} included optical observations of the cluster
region centred on 3C\,353 taken with the University of Hawaii 2.2\,m
telescope in their analysis of Zw\,1718.1-0108.  They identified three
additional massive galaxies, none of which reside in the northern
sub-cluster.  Deeper optical observations would help to establish
whether there are any galaxies associated with the northern
sub-cluster and whether 3C\,353 is the dominant member of this
sub-cluster despite its position at the edge.

\section{Summary and Conclusions}

Our results can be summarized as follows:

\begin{itemize}

\item By fitting an inverse-Compton model to the lobes, we found the
East lobe to be electron dominated and the West lobe to be consistent
(within the large errors) with equipartition.

\item We determined that a variation in the electron spectron cannot
account for the varying X-ray and radio emission alone, but that a
change in the magnetic field strength across the lobes is required.

\item We have obtained a good X-ray spectrum of the nucleus of
3C\,353.  Both the X-ray and optical properties of this source are
ambiguous but it appears to lie in a region of parameter space in
between those normally occupied by narrow-line and low-excitation
radio galaxies.

\item We have detected an X-ray counterpart for the East hotspot
offset by $4.0 \pm 0.5$\,kpc.

\item The northern and southern sub-clusters were found to be
isothermal with a temperature difference of $\sim 1$\,keV supporting a
model in which they are two originally separate components undergoing
a merger with no evidence for a violent interaction.

\end{itemize}

\section*{Acknowledgements}

JLG thanks the STFC for a research studentship. MJH acknowledges
generous financial support from the Royal Society.  We also thank Jun
Kataoka and the anonymous referee for helpful advice and comments. The
National Radio Astronomy Observatory is a facility of the National
Science Foundation operated under cooperative agreement by Associated
Universities, Inc. We thank the staff of the GMRT for their help with
the observations with that telescope: GMRT is run by the National
Centre for Radio Astrophysics of the Tata Institute of Fundamental
Research, India. This work is partly based on observations obtained
with {\it XMM-Newton}, an ESA science mission with instruments and
contributions directly funded by ESA Member States and NASA.  Basic
research in radio astronomy at the Naval Research Laboratory is
supported by 6.1 basic research.

\label{lastpage}

\end{document}